\documentclass[aps,pr,reprint,superscriptaddress,longbibliography]{revtex4-1}

\usepackage[separate-uncertainty=true,
    multi-part-units=single,
    per-mode=symbol,
    group-separator={,},
    range-phrase=--,
    range-units=brackets]{siunitx}
\usepackage{graphicx}
\graphicspath{{figures/}}

\setcitestyle{numbers,comma,square}

\usepackage{amsmath}

\newcommand{\mgml}{\milli\gram\per\milli\liter}
\renewcommand{\vec}[1]{\mathbf{#1}}
\newcommand{\abs}[1]{\left\vert #1 \right\vert}

\begin{document}

\title{Quantitative differentiation of protein aggregates from other subvisible particles in viscous mixtures through holographic characterization}

\author{Annemarie Winters}
\author{Fook Chiong Cheong}
\author{Mary Ann Odete}
\author{Juliana Lumer}
\author{David B. Ruffner}
\affiliation{Spheryx, Inc., 330 E.\ 38th St., New York, NY 10016, USA}

\author{Kimberly I. Mishra}
\affiliation{Spheryx, Inc., 330 E.\ 38th St., New York, NY 10016, USA}
\affiliation{Department of Physics and Center for Soft Matter Research, 
  New York University, New York, NY 10003, USA}

\author{David G. Grier}
\affiliation{Department of Physics and Center for Soft Matter Research, 
  New York University, New York, NY 10003, USA}

\author{Laura A. Philips}
\affiliation{Spheryx, Inc., 330 E.\ 38th St., New York, NY 10016, USA}

\begin{abstract}
  We demonstrate the use of holographic video microscopy to
  detect individual subvisible particles dispersed in biopharmaceutical
  formulations and to differentiate them based on material characteristics
  measured from their holograms.
  The result of holographic analysis is a precise and accurate measurement
  of the concentrations and size distributions of multiple classes of
  subvisible contaminants dispersed in the same product simultaneously.
  We demonstrate this analytical technique through measurements on model systems
  consisting of human IgG aggregates
  in the presence of common contaminants such as
  silicone oil emulsion droplets and fatty acids. 
  Holographic video microscopy also
  clearly identifies metal particles and air bubbles.
  Being able to differentiate and characterize the individual
  components of such heterogeneous dispersions provides a basis for 
  tracking other factors that influence
  the stability of protein formulations
  including handling and
  degradation of surfactant and other excipients.
\end{abstract}

\maketitle

\section{Introduction}
\label{sec:introduction}

Ensuring the safety and efficacy of protein-based
pharmaceuticals benefits from %requires 
methods to detect subvisible
particulate contaminants, to differentiate them by
composition, and to measure the concentrations of
each population of particles in dispersion
\cite{ratanji2014immunogenicity}.
We previously have demonstrated 
that holographic video microscopy (HVM)
can detect individual contaminant particles ranging in size
from \SI{500}{\nm} to \SI{10}{\um} and can differentiate
subvisible protein aggregates from silicone oil 
emulsion droplets
on the basis of their differing refractive indexes \cite{wang16,kasimbeg2019holographic}.
Here, we demonstrate that HVM can
detect, differentiate and identify multiple distinct
populations of subvisible particles when
they are present simultaneously 
in complex heterogeneous
dispersions, including the most common categories of subvisible contaminants that are introduced into biopharmaceutical products
during the various stages of development, manufacturing and use.
These include subvisible protein aggregates 
in combination with oil droplets
\cite{felsovalyi2012silicone,shah2017evaluation},
degradants of surfactants \cite{martos2017trends},
metal particles \cite{tyagi2009igg} and air bubbles
\cite{scherer2012issues}.
All such particle types are indistinguishable to
conventional particle characterization technologies
including microflow imaging (MFI) and HIAC \cite{scherer2012issues}.
They are rapidly and reliably detected, differentiated
and quantitated by HVM.

Differentiating colloidal particles by refractive index 
is a unique capability of HVM
relative to other particle characterization techniques \cite{xu01,wang16}.
Single-particle HVM measurements proceed 
rapidly enough to build
up statistics on tens of thousands of particles in twenty minutes \cite{cheong09,krishnatreya14,yevick14,hannel18}.
These results then yield
the concentrations and size distributions of each population of particles in a complex mixture \cite{wang16,philips17}.

\begin{figure}
    \centering
    \includegraphics[width=0.9\columnwidth]{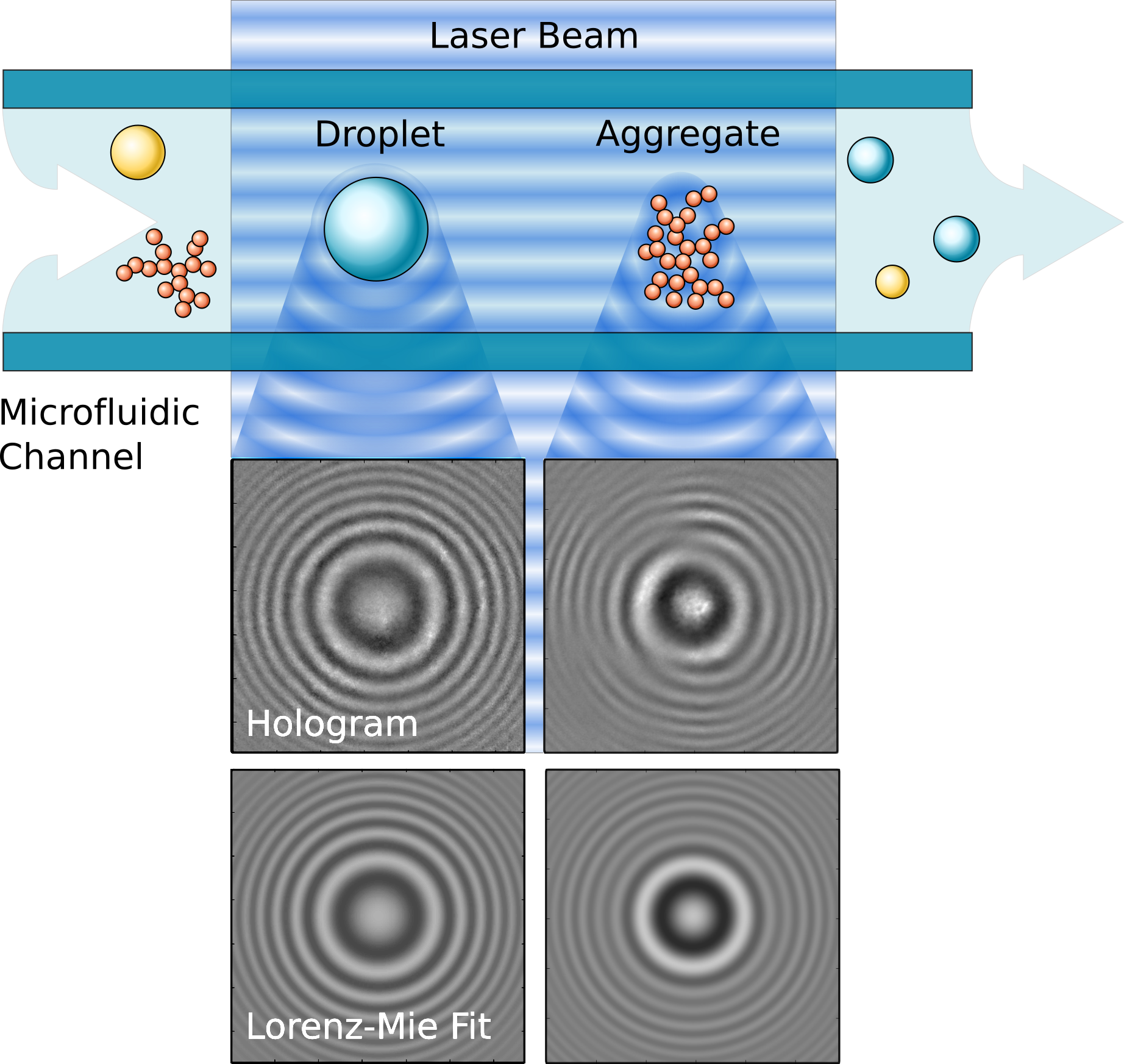}
    \caption{Principle of holographic particle characterization. Subvisible particles flow down a microfluidic channel through a collimated laser beam. Light scattered by a particle
    interferes with the remainder of the
    beam to create a hologram of the
    particle that is magnified by a
    microscope (not shown) and recorded
    with a video camera. Each hologram is
    fit pixel-by-pixel to a
    generative model derived from the
    Lorenz-Mie theory of light scattering
    to obtain that particle's
    effective diameter and refractive
    index.
    }
    \label{fig:schematic}
  \end{figure}

Real-world biopharmaceutical products not only play
host to a wide variety of contaminant particles,
but also have widely varying physical characteristics, most notably variations in viscosity that
can pose challenges to standard measurement techniques.
Previous HVM studies of protein aggregation
have been performed in water with a
viscosity around \SI{1}{\centi P}
\cite{wang16,kasimbeg2019holographic}.
We establish through measurements on NIST-traceable
colloidal standards that HVM also yields correct results for the
diameter and refractive index
of subvisible colloidal particles
across the commercially relevant range of viscosities,
up to \SI{20}{\centi P}.
Titration studies also show that
HVM provides consistent results for
concentration across this range.

Changes in medium composition can influence
the refractive index of the medium.
Medium refractive index, however, does not influence
holographic characterization of compact objects such as
oil droplets, metal particles, and air bubbles, whose 
intrinsic light-scattering properties are
not influenced by the medium
\cite{cheong11,odete20}.
Changes in the medium's refractive index do change
the signature of porous objects such
as protein aggregates, whose measured refractive indexes
track changes in the index of the medium. Such changes also
can be used to distinguish protein aggregates from other,
compact homogeneous contaminants.

Some contaminants, such as the fatty-acid breakdown products of
standard surfactants, have quite similar optical characteristics
to protein aggregates. 
Their presence nonetheless can be inferred
from holographic characterization measurements through their influence
on the distribution of detected particle properties.

\section{Methods and Materials}
\label{sec:methodsmaterials}

\subsection{Holographic Video Microscopy}
\label{sec:characterization}

Holographic video microscopy measurements are performed
with xSight (Spheryx, Inc.), which is
a turn-key commercial implementation of the holographic
characterization instrument described in Ref.~[\onlinecite{wang16}].
The measurement principle is presented in Fig.~\ref{fig:schematic}.
Characterizing a sample involves pipetting
a \SI{30}{\micro\liter} aliquot into the
reservoir of a disposable xCell microfluidic sample chip.
xSight engages a vacuum pump with the chip to pull
the sample in a pressure-driven Poiseuille flow through the
xCell's observation volume, where it is illuminated by a collimated
laser beam at a vacuum wavelength of \SI{447}{\nm}.
Colloidal particles in the fluid stream scatter some of this
illumination to the focal plane of an optical microscope, where
it interferes with the rest of the beam.  The microscope magnifies
the resulting interference pattern and relays it to a video camera that records
its intensity.

Each snapshot recorded by xSight's camera constitutes 
a hologram of
the particles in the xCell's observation volume and therefore
encodes information about the particles' three-dimensional
positions, their diameters and their refractive indexes.
This information is extracted by fitting each single-particle
hologram to a generative model \cite{lee07a} based on the 
Lorenz-Mie theory of light scattering \cite{bohren83,mishchenko02}.
Details of the analysis are presented in the Appendix.

Each fit yields the particle's diameter, $d_p$, with a precision of \SI{+-5}{\nm} and its refractive index, $n_p$,
to within \num{+-0.003}
\cite{krishnatreya14}.
Holographically measured tracking data are
used to follow each particle's motion through the sample volume,
both to validate the flow profile and also to provide multiple
independent measurements of each particle's properties.

Holographic measurements of particle sizes
and refractive index are parameterized
by the wavelength of light,
the magnification of the microscope, and
the refractive index, $n_m$, of the fluid medium, the last of which
can be obtained at part-per-thousand precision with an Abbe
refractometer.
No additional calibration measurements are required.
Instrumental precision and accuracy are
validated by measurements on NIST standard particles, as described
in Ref.~[\onlinecite{wang16}].

Holographic video microscopy is most effective for subvisible
particles ranging in diameter from \SI{500}{\nm} to \SI{10}{\um} and
for concentrations ranging from \SI{e3}{particles\per\milli\liter}
to \SI{e7}{particles\per\milli\liter} \cite{wang16}.
A twenty-minute measurement
inspects all of the particles in \SI{3}{\micro\liter} of the sample,
yielding estimates for particle concentrations whose precision is
limited on the low end by counting statistics
and on the high end by occlusion 
\cite{kasimbeg2019holographic}.

A particle's refractive index is determined by its composition
\cite{lee07a,yevick14} and thus provides a basis for differentiating subvisible contaminants of different composition. HVM is unique among particle characterization techniques
in its ability to provide this information \cite{xu01}.

\subsection{Preparation of Dispersions of Subvisible Particles}
\label{sec:samplepreparation}

The model multicomponent 
colloidal dispersions 
analyzed in this study
are created by mixing stock 
solutions, emulsions and single-component
colloidal dispersions in clean 
\SI{12}{\milli\liter} vials,
inverting \SI{10}{times} and then vortexing for 
\SI{10}{\second}.
Although vortexing can introduce
air bubbles in some samples,
any such bubbles would not compromise
HVM analysis of particle properties because
HVM can distinguish air bubbles from other
particles.
Freshly 
mixed samples are transferred immediately into
the \SI{30}{\micro\liter} reservoir of a fresh
xCell channel for analysis.
All raw materials are used as delivered by
the supplier.

\subsubsection{Immunoglobulin G (IgG) aggregates}
\label{sec:igg}

A stock solution of human IgG 
is prepared by dissolving 
lyophilized low-endotoxin IgG 
(Molecular Innovations, catalog no.\ HU-GF-ED)
in filtered DI water at 
room temperature
to a concentration of \SI{16}{\mgml}.
IgG readily forms subvisible aggregates under
these conditions, as is confirmed
by HVM measurements.

\subsubsection{Polystyrene standard spheres}
\label{sec:ps}

NIST-traceable polystyrene spheres 
(Bangs Laboratories, catalog no.\ NT16N) with a
nominal diameter of $d_p = \SI{1.54}{\um}$
are dispersed in DI water at a concentration
of \SI{4e6}{particles\per\milli\liter}.
HVM confirms a
population-mean diameter of
$d_p = \SI{1.54(5)}{\um}$
and a refractive index of
$n_p = \num{1.603(3)}$, which is consistent
with expectations for polystyrene \cite{sultanova2009dispersion}.

\subsubsection{Silica standard spheres}
\label{sec:silica}

NIST-traceable silica spheres (Bangs Laboratories,
catalog no.\ SS04N) with a nominal diameter of
$d_p = \SI{2.2}{\um}$ are dispersed in DI water
at a concentration of \SI{4e6}{particles\per\milli\liter}.
HVM confirms a population-mean diameter
of $d_p = \SI{2.20(5)}{\um}$ and a refractive
index of $n_p = \num{1.424(5)}$, which is 
consistent with expectations for silica.

%Whenever I say DI water, I mean filtered DI water that was filtered with a MILLEX GP 0.22μm Filter Unit. 

%IgG - PS controls:

%No sugar: 0.5mL 1.0mg/mL IgG + 0.5mL 4e6 PS (1.54um) + 2mL DI % Water
%½ sugar: 0.5mL 1.0mg/mL IgG + 0.5mL 4e6 PS (1.54um) + 1mL 66% Sucrose + 1mL DI Water
%Full sugar: 0.5mL 1.0mg/mL IgG + 0.5mL 4e6 PS (1.54um) + 2mL 66% Sucrose

%4e6 PS solution: made a 4e7 solution first: 80.3uL source 1.54um PS solution + 9.92mL DI water. Then diluted by 10 to get 4e6 solution. 

%Prepped sample by adding all components, then inverting ten times, then vortexing for ten seconds before loading into xCell. 
\subsubsection{Silicone oil emulsion}
\label{sec:siliconeemulsion}

Silicone oil (Sigma-Aldrich catalog no.\ 378399,
CAS no.\ 63148-62-9, 
MDL no.\ MFCD00132673,
\SI{1}{\centi P}) is added to DI water at
\SI{26}{\mgml}.
The sample is shaken vigorously by hand to
disperse the silicone oil as emulsion droplets.
HVM confirms that the resulting droplets
have a broad distribution of diameters
but a very narrow distribution of
refractive indexes centered at 
$n_p = \num{1.410(3)}$.

\subsubsection{Oleic acid dispersion}
\label{sec:oleicacid}

Oleic acid ($\ge\SI{90}{\percent}$, Sigma-Aldrich
catalog no.\ 364525, CAS no.\ 112-80-1, 
MDL no.\ MFCD00064242) 
is dissolved in methanol 
($\ge\SI{99}{\percent}$,
Sigma-Aldrich catalog no.\ M3641, 
CAS no.\ 67-56-1, MDL number MFCD00004595)
at \SI{0.2}{\percent} by volume
and then is precipitated as droplets by
$10\times$ dilution in DI water.

\subsubsection{Stearic acid dispersion}
\label{sec:stearicacid}

Stearic acid ($\ge \SI{98}{\percent}$, 
Alfa Aesar catalog no.\ A12244, CAS no.\ 57-11-4)
is dissolved in methanol at a concentration of
\SI{9}{\mgml}.
Aggregates of stearic acid particles are
precipitated from this solution by $10\times$
dilution in DI water. This dispersion then is
further diluted by a factor of \num{100}
in DI water.

\subsubsection{Tungsten particles}
\label{sec:tungsten}

Tungsten particles 
(US Research Nanomaterials catalog no.\ US5014)
with a nominal diameter of $d_p = \SI{300}{\nm}$
are added to DI water at \SI{13.6}{\mgml}
and are dispersed by vortexing for
\SI{15}{\second}.

\subsubsection{Air bubbles}
\label{sec:airbubbles}

Micrometer-scale 
air bubbles are introduced directly into
the reservoir of an xCell by rapidly ejecting
an aqueous solution of polysorbate 20 (PS20)
and sucrose from a 31G insulin syringe (Sure Comfort U-100).
The solution uses \SI{1}{\mgml} PS20 
(Alfa Aesar catalog no.\ L15029, CAS no.\ 9005-64-5)
as a foaming agent and \SI{64}{\percent wt}
sucrose 
(Carolina Biological Supply 
catalog no.\ 892860, CAS no.\ 57-50-1,
MDL no.\ MFCD00006626)
to increase the viscosity to 
roughly \SI{20}{\centi P}
\cite{telis2007viscosity}.

\subsubsection{Tuning the dispersion's 
refractive index}
\label{sec:tuningmedium}

The refractive index, $n_m$, of the aqueous
medium is adjusted by adding sucrose (Carolina Biological Supply,
catalog no.\ 892860, CAS no.\ 57-50-1, MDL no.\ MFCD00006626)
or glycerol (Sigma-Aldrich, catalog no.\ G5516, 
CAS no.\ 56-81-5, MDL no.\ MFCD00004722).
In each case the dispersion to be studied
is prepared as a stock sample at
$10\times$ the desired particle concentration.
The same stock dispersion then is diluted to
\SI{10}{v/v\percent} for each preparation
in an aqueous solution of sucrose or
glycerol whose concentration is chosen
to provide a desired value of $n_m$
\cite{reis2010refractive}.
The actual value of $n_m$
is determined with an Abbe refractometer
(Edmund Optics).
% {\color{red}[XXX what refractometer???].}
This value also is used to confirm
the concentration of sucrose or glycerol
in solution \cite{reis2010refractive}.
Diluting a stock dispersion by a fixed
proportion ensures that the same 
concentration of particles is present
in each dispersion across the range
of refractive indexes studied.

Sucrose and glycerol both increase the
viscosity of aqueous solutions. 
xSight accommodates samples with viscosities
ranging from \SI{1}{\centi P}
to \SI{25}{\centi P}, which encompasses
the range of dispersion viscosities used
in this study.
The precise value of the viscosity
is not required for successful HVM
measurements and is estimated from
the solution's concentration
\cite{segur1951viscosity,telis2007viscosity}.

\section{Results}
\label{sec:results}

\subsection{Detection and differentiation
of subvisible contaminant particles in a complex
heterogeneous sample}

\begin{figure*}
    \centering
    \includegraphics[width=0.75\textwidth]{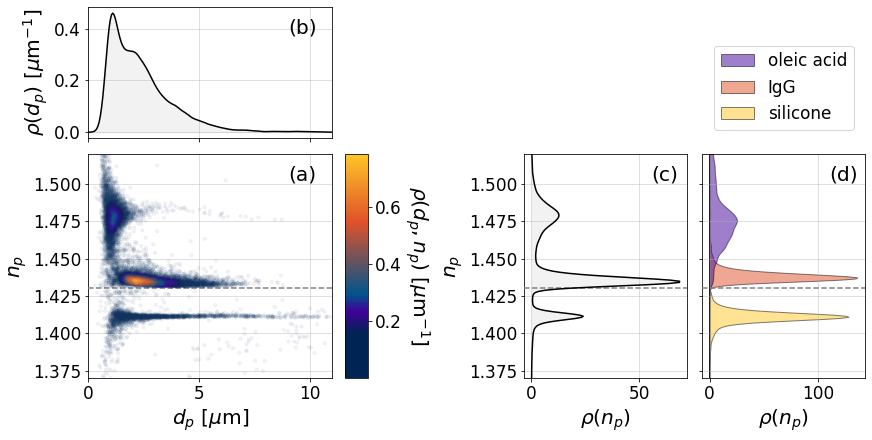}
    \caption{Holographic characterization of contaminant particles
    in a solution of human IgG. Suspended particles include
    IgG aggregates, silicone oil droplets and
    oleic acid droplets. The aqueous medium includes dissolved sucrose
    that raises the medium's refractive index to $n_m = \num{1.430(1)}$ and increases the medium's viscosity.
    (a) Scatter plot in which each point represents
    the diameter, $d_p$, and refractive index, $n_p$, 
    of one detected particle and is colored by the density
    of measurements, $\rho(d_p, n_p)$. 
    (b) The projected size distribution,
    $\rho(d_p)$, yields the total concentration of detected subvisible
    particles.
    (c) The projected distribution of refractive indexes, $\rho(n_p)$,
    distinguishes particles by composition.
    (d) Superimposed 
    projected refractive index distributions of three control
    samples of oleic acid droplets, IgG aggregrates and
    silicone oil droplets.}
    \label{fig:3mix}
\end{figure*}

Figure~\ref{fig:3mix}(a) shows HVM results for a
sample containing a mixture of protein aggregates,
silicone oil emulsion droplets and droplets of oleic
acid.
Each of the \num{18892}
discrete points in the scatter plot represents
the diameter, $d_p$, and refractive index, $n_p$,
of a single particle that was detected in 
\SI{3}{\micro\liter}
of the sample.
The dots fall into clusters that represent
different populations of particles.
The density of measurements, $\rho(d_p, n_p)$,
therefore offers insights into the composition of
the sample.
Each point in Fig.~\ref{fig:3mix}(a) is colored
by the density of particles, $\rho(d_p, n_p)$, as indicated by the color bar.

The size distribution of all of the
particles in the sample, $\rho(d_p)$, is
presented in Fig.~\ref{fig:3mix}(b).
It combines information from all three populations
of particles and does not provide a basis for
distinguishing among them.
This projected size distribution shows that there are
many more small particles than large in this sample, down
to the \SI{500}{\nm} lower detection limit of the
instrument.
The decrease in observed particle concentration
for the smallest particles reflects the loss of
detection sensitivity near the
instrumental limit.

The multicomponent nature of the sample is clearly evident
in the projected distribution of single-particle 
refractive indexes, $\rho(n_p)$, which is plotted in Fig.~\ref{fig:3mix}(c).
Broadly speaking, this plot reveals three populations of particles, each represented as a distinct peak in
$\rho(n_p)$.
The nature of each population is revealed in 
the joint distribution of single-particle
sizes and refractive indexes
in Fig.~\ref{fig:3mix}(a).
The lowest-index population of particles appears as
an extended horizontal stripe in $\rho(d_p, n_p)$,
which means that the detected particles have a wide
range of sizes, but a narrow distribution of
refractive indexes.
Such horizontal stripes are characteristic
of emulsions whose droplets all have the same
composition and therefore have similar refractive indexes.

The low-index population has a refractive
index of $n_p = \num{1.410(3)}$, which is below that of the 
medium, as indicated by the horizontal dashed line
in Fig.~\ref{fig:3mix}.
The two higher-index populations have
refractive indexes higher than that of the medium.
These results show that
HVM works equally well for particle indexes
above and below the index of the medium,
even when low- and high-index particles appear
in the same sample.

Particles in the middle population have 
a broad range of sizes and
refractive indexes very close to
that of the medium.
Within this population, 
smaller particles tend to have higher
refractive indexes.
These trends are characteristic of porous
particles whose pores are filled with the medium
\cite{wang16,wang16a,kasimbeg2019holographic,fung2019assessing}.
Their measured refractive indexes
are intermediate between that of the particle's
matrix and that of the medium \cite{cheong11,odete20}.
HVM has been demonstrated to yield accurate
characterization data for such
inhomogeneous and aspherical particles 
\cite{hannel15,wang16,wang16a,fung2019assessing,odete20}
through the effective medium theory of light
scattering \cite{markel16,cheong11}.
In this case, the  population
of particles near the medium refractive index is naturally identified with
protein aggregates, as distinct from emulsion
droplets.

The highest-index peak is
centered symmetrically around $n_p = \num{1.475(10)}$,
which is slightly lower than the 
previously published value of \num{1.489} 
for oleic acid
at the imaging wavelength \cite{yanina2018refractive}.
Aerosol droplets of oleic acid precipitated
from alcohol are reported to remain in fluid
state \cite{secker2000light}.
Smaller droplets, however, are found to aggregate into
irregular clusters without coalescing
\cite{katrib2005ozonolysis}.
This would account for the slightly low
value of the measured refractive index
and the comparatively broad distribution of
refractive index values 
\cite{cheong11,wang16,wang16a,fung2019assessing}.

This interpretation of the HVM data is supported
by characterization measurements performed on the
component single-population samples independently.
These measurements yield the refractive index
distributions plotted in Fig.~\ref{fig:3mix}(d)
whose superimposed peaks correspond with those
in the mixed sample.
Slight differences in the peak shapes may
reflect interactions among particles
from different populations in the mixed sample.

\subsection{Inorganic particles: air bubbles and metal particles}
\label{sec:inorganic}

\begin{figure*}
    \centering
    \includegraphics[width=0.75\textwidth]{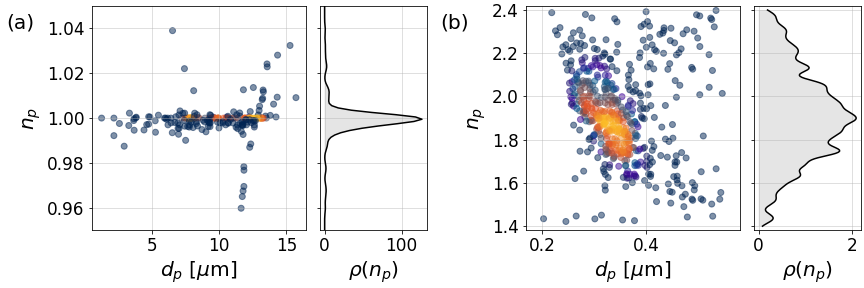}
    \caption{Joint distribution of the diameter, $d_p$,
    and refractive index, $n_p$, of (a) air bubbles and (b) tungsten spheres, together with the
    projected distributions of refractive indexes,
    $\rho(n_p)$. Air bubbles have refractive indexes
    very tightly clustered around $n_p = \num{1.0}$.
    Monodisperse tungsten spheres display a comparatively small range of diameters, and very high refractive index values.}
    \label{fig:inorganic}
\end{figure*}

 Air bubbles tend to form in viscous formulations
subjected to agitation or
fast ejection from syringes \cite{corvari2015subvisible,randolph2015not}.
Figure~\ref{fig:inorganic}(a) shows that
HVM can distinguish bubbles 
from dispersed particles and droplets by their refractive 
index, $n_p = \num{1.00}$, which is the refractive index
for most gases, including air.
This natural basis for identifying bubbles
is an advantage of HVM relative to techniques
such as HIAC and FlowCAM that cannot easily differentiate
subvisible bubbles from other suspended and 
dispersed species \cite{werk2014effect,hawe2015subvisible}.

Metal fragments similarly have a clear HVM signature,
as can be seen in the data for tungsten particles 
presented in Fig.~\ref{fig:inorganic}(b).
Metal particles tend to have refractive indexes
that are substantially higher than organic matter.
Tungsten and other metal particles can contaminate 
pharmaceutical products at all stages of 
manufacturing and can influence product stability, efficacy 
and safety \cite{vanbeers2012immunogenicity}.
Metal particles also are not readily distinguished from other
dispersed species by standard particle characterization
techniques \cite{werk2014effect,hawe2015subvisible}.
The particles reported in Fig.~\ref{fig:inorganic}(b)
have a mean refractive index of $n_p = \num{1.85(9)}$
which greatly exceeds values for protein aggregates,
silicone oil, or degradants such
as breakdown products of surfactants.
Refractive index therefore provides a natural
basis for identifying metal particles in multicomponent
dispersions.

\subsection{Differentiating subvisible spheres: air bubbles and silicone oil droplets}
\label{sec:bubbleoil}

\begin{figure}
    \centering
    \includegraphics[width=\columnwidth]{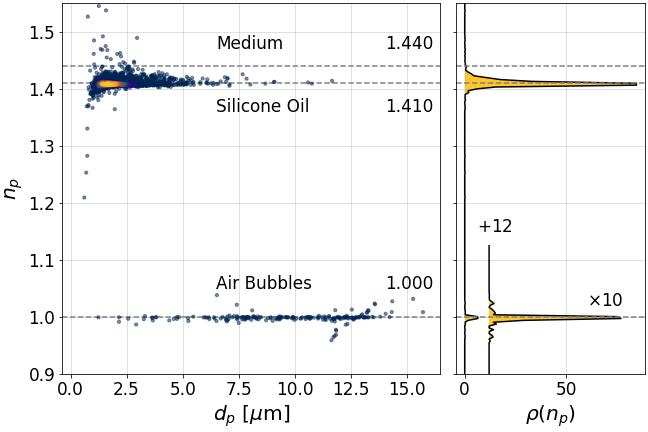}
    \caption{Holographic analysis of silicone oil droplets
    and air bubbles dispersed simultaneously in a viscous medium.
    Air bubbles have a refractive index of 
    \num{1.000}. Silicone oil droplets have a refractive
    index of \num{1.410}. The aqueous medium has
    a refractive index of $n_m = \num{1.440}$.
    The distribution of refractive index values,
    $\rho(n_p)$, shows two clearly resolved peaks.
    There being far fewer bubbles than droplets in
    this sample, the peak around $n_p = \num{1.000}$
    is multiplied by \num{10} and displaced
    by \num{12} for clarity.}
    \label{fig:bubbleoil}
\end{figure}

Being perfectly spherical,
air bubbles and silicone oil droplets
are readily differentiated from irregular
aggregates with conventional imaging
techniques, but can be
challenging to distinguish from each other.
The data in Fig.~\ref{fig:bubbleoil} demonstrate
that HVM unambiguously differentiates air bubbles from
silicone oil droplets on the basis of refractive index
when both appear in the same sample.
The refractive index of silicone oil
droplets depends on the chemical composition
of the oil.
Air bubbles all have the same
refractive index, $n_p = \num{1.000}$.

The peak placement 
in the measured distribution, $\rho(n_p)$,
further validates the precision and accuracy
of HVM for single-particle measurement.
The refractive indexes of each detected particle
is discovered by fitting 
rather than being assumed \emph{a priori}.
The peak associated with the air bubbles
therefore provides an unambiguous 
reference point.

\subsection{Concentration measurements}
\label{sec:concentration}

\begin{figure}
    \centering
    \includegraphics[width=0.8\columnwidth]{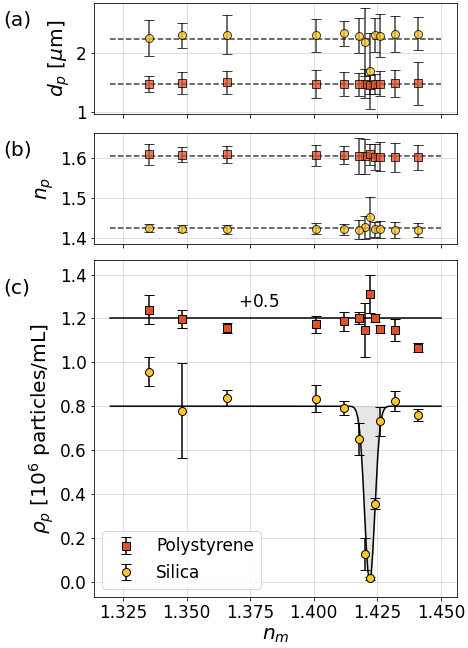}
    \caption{
    (a) Population-averaged values for the diameters, 
    $d_p$, of polystyrene spheres
    (orange squares)
    and silica spheres (yellow circles) dispersed
    in water-glycerol mixtures of varying refractive
    index, $n_m$.
    Measured diameters are independent of $n_m$.
    (b) The particles' refractive indexes, $n_p$, 
    similarly are independent of $n_m$.
    (c) Measured concentrations, $\rho_p$, of 
    polystyrene and silica spheres as a function of $n_m$.
    The concentration of each population of particles is
    reported consistently and is independent of $n_m$
    except for a region, $n_p = n_m \pm \num{0.002}$ 
    in which
    silica spheres are index-matched to the medium and
    therefore are not detectable.
    Concentration data for polystyrene spheres are
    offset upward by \SI{5e5}{particles\per\milli\liter}
    for clarity.
   }
    \label{fig:concentration}
\end{figure}

In addition to detecting and differentiating different
types of particles in a multicomponent sample, HVM
also accurately measures the concentration of each of the
populations.
We demonstrate this capability with a series of samples
composed of an aqueous dispersion of
\SI{1.5}{\um}-diameter polystyrene spheres
and \SI{2.2}{\um}-diameter silica spheres
diluted to an overall concentration of 
\SI{1.5e6}{particles\per\milli\liter}
by the addition of glycerol-water solutions.
Depending on the final concentration of glycerol,
the medium's refractive index ranges from $n_m = \num{1.34}$ 
for pure water to $n_m = \num{1.44}$.
Over the same range, the medium's viscosity ranges from
\SI{0.89}{\centi P} to nearly \SI{20}{\centi P}.
The particles' diameters and refractive indexes,
however, should remain constant throughout,
as should the concentrations of the two populations.

HVM readily distinguishes the two population of spheres 
both by diameter and also
by refractive index, as can be appreciated from
Fig.~\ref{fig:concentration}(a) and \ref{fig:concentration}(b).
Results for $d_p$ and $n_p$, moreover, are 
independent of the medium's refractive
index and viscosity, as expected.

Dividing the number of particles detected
in each population by the volume of fluid
analyzed yields that population's concentration.
Figure~\ref{fig:concentration}(c) shows the
detected concentrations of the two populations of
spheres over the same range of medium compositions.
These concentrations also are independent
of the medium's composition, except for a very
narrow range of refractive-index values centered
around $n_m = \num{1.422}$. 
In this window, the silica spheres
are index-matched to the medium, 
and so cannot be detected and counted
by optical means.
This effect does not influence the measured concentration
of polystyrene spheres codispersed in the same medium
because polystyrene's refractive index, 
$n_p = \num{1.601}$, differs substantially from that of the
medium.

Index matching affects concentration measurements over
a remarkably narrow range of refractive indexes.
The solid curve in Fig.~\ref{fig:concentration}(c)
is a Gaussian with width $\Delta n_p = \num{0.002}$.
HVM reliably detects and reports the concentration of
particles whose refractive indexes differ by more than
$\Delta n_p$ from $n_m$.
The reported concentration values, moreover, do not
depend on the physical properties of the medium,
including chemical composition, viscosity and
refractive index.

\subsection{Influence of the medium}
\label{sec:medium}

\begin{figure}
    \centering
    \includegraphics[width=0.9\columnwidth]{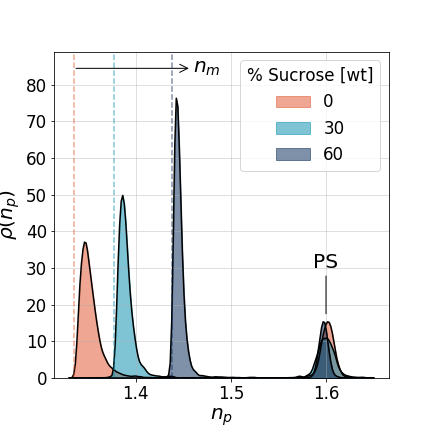}
    \caption{Influence of added sucrose on the holographically
    measured refractive index distribution of IgG aggregates.
    Codispersed polystyrene beads (PS) serve as a reference.}
    \label{fig:sucrose}
\end{figure}

Changes in the properties of the medium due
to added excipients can influence the results
of HVM measurements, most notably for porous
particles whose pores are perfused with the medium
\cite{cheong11,odete20}.
The effective refractive index reported by HVM
for such particles is intermediate between
the refractive index of the medium and the
refractive index of the porous particles'
matrix material.
This mechanism was invoked in the discussion
of protein aggregates' properties presented in
Fig.~\ref{fig:3mix}.
The data in Fig.~\ref{fig:sucrose} show this
mechanism in action.
These results are obtained for mixtures of
IgG aggregates and polystyrene spheres in
solutions with \SI{0}{\percent}, \SI{30}{\percent}
and \SI{60}{\percent} sucrose by weight.
These solutions have refractive indexes of
$n_m = \num{1.335}$, \num{1.377} and \num{1.438},
respectively and viscosities of
\SI{1}{\centi P}, \SI{3}{\centi P}
and \SI{58}{\centi P}, respectively.

The polystyrene particles in these dispersions
yield refractive indexes consistent with
$n_p = \num{1.610(5)}$, independent of 
medium composition and in agreement with the
results from Fig.~\ref{fig:concentration}(b).
Polystyrene spheres are non-porous and hydrophobic, 
which means that the
medium should not influence their optical properties,
as observed.

The mean refractive index of protein aggregates
tracks the refractive index of the medium.
The distribution of refractive index values
furthermore narrows as the refractive index of
the medium increases toward the refractive index of
protein.
Both of these trends are consistent with
predictions of the Maxwell Garnett effective medium
theory for light-scattering by inhomogeneous media
\cite{markel16,cheong11,wang16a,fung2019assessing,odete20}.

\subsection{Influence of handling}
\label{sec:handling}

\begin{figure*}
    \centering
    \includegraphics[width=0.75\textwidth]{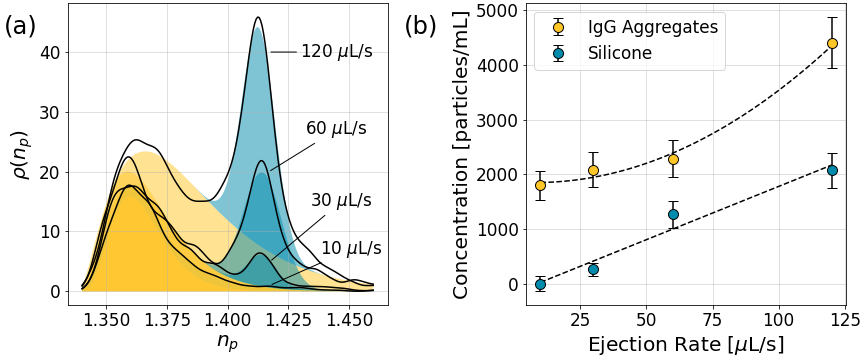}
    \caption{Influence of syringe ejection rate on particle
    concentrations. (a) Measured particle refractive index
    distributions, $\rho(n_p)$, for four different 
    ejection rates. Dark (cyan) shaded regions 
    represent the symmetric
    Gaussian distribution expected for silicone oil droplets.
    Lighter (yellow) shaded regions represent an
    asymmetric Gamma distribution for protein aggregates.
    Their sum is a model for the total measured distribution
    and serve to identify each population in the sample.
    (b) Integrated concentrations of protein aggregates
    and silicone oil emulsion droplets obtained from the
    data in (a) as a function of ejection rate.}
    \label{fig:flowrate}
\end{figure*}

Handling conditions can change
the concentration and composition of the
particles in a protein solution.
The data in Fig.~\ref{fig:flowrate}(a)
show the refractive index distribution,
$\rho(n_p)$, for a solution of human IgG
in water 
($n_m = \num{1.340}$) 
as a function of ejection rate from a 
\SI{1}{\milli\liter} syringe through a 31G needle.
The syringes used for this study
(Beckton-Dickinson, BD Safety-Glide\texttrademark\ \SI{1}{\milli\liter} insulin syringe with 
BD Ultra-Fine\texttrademark\ needle)
are lubricated with
silicone oil and are known to release
oil droplets \cite{melo2019release}.

The distribution shows two populations of particles, one
peaked asymmetrically around
$n_p = \num{1.36}$ and the other 
centered symmetrically and more narrowly 
around $n_p = \num{1.41}$.
We interpret the former as representing
a population of protein aggregates, and
the latter as arising from a population
of silicone oil droplets.

The concentration and distribution of
particle properties clearly changes as
the ejection rate is increased from
\SI{10}{\micro\liter\per\second} up
to \SI{120}{\micro\liter\per\second}.
To quantify these trends, we fit 
$\rho(n_p)$ to the sum of a 
symmetric Gaussian distribution representing
the silicone oil droplets
and an asymmetric Gamma distribution representing
protein aggregates.
These fits appear as shaded
regions in Fig.~\ref{fig:flowrate}(a),
with lighter (yellow) shading corresponding
to the fit for protein aggregates and
darker (cyan) shading corresponding to
silicone oil droplets.

The areas under these curves correspond to
the numbers of particles of each type observed
in \SI{0.5}{\micro\liter} of the sample.
The associated concentrations are plotted
as a function of ejection rate in 
Fig.~\ref{fig:flowrate}(b).
Essentially no silicone oil droplets
appear in the sample ejected at low
flow rates.
Faster flows elute more silicone oil droplets,
with the concentration rising to 
\SI{2000}{droplets\per\milli\liter}
at an ejection rate of \SI{120}{\micro\liter\per\second}.
Interestingly, the concentration of protein aggregates
doubles over the same range of ejection rates.
This trend is only visible because HVM
differentiates the two types of particles.
The data in Fig.~\ref{fig:flowrate} therefore
highlight the value of HVM for detecting and
interpreting changes in protein solutions
induced by handling, in this case flow-induced
changes.

\begin{figure*}
    \centering
    \includegraphics[width=0.75\textwidth]{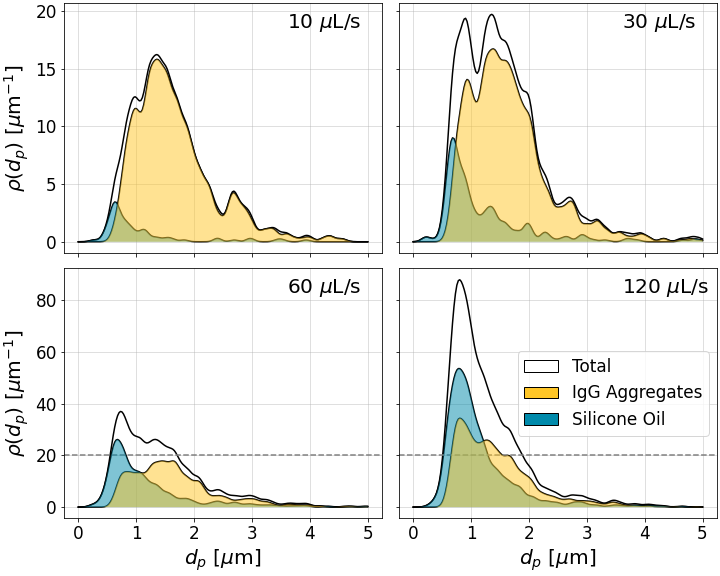}
    \caption{Influence of syringe ejection rate on
    the distribution of particle diameters, $\rho(d_p)$,
    in the samples presented in Fig.~\ref{fig:flowrate}.
    Curves show the total distribution of all detected
    particles as well as distributions for IgG aggregates
    and silicone oil emulsion droplets identified
    on the basis of refractive index.
    }
    \label{fig:speeddiameter}
\end{figure*}

Figure~\ref{fig:speeddiameter} shows the distribution
of particle diameters, $\rho(d_p)$ for the same samples,
presented as a function of ejection rate.
For each sample, curves show the total distribution
for all particles in the sample as well as separate
distributions for the IgG aggregates and silicone
oil droplets that were differentiated by
refractive index.
Consistent with the conclusions drawn from the
refractive-index data in Fig.~\ref{fig:flowrate},
increased elution rate increases the concentration
of aggregates and oil droplets alike at all sizes.

\section{Discussion}
\label{sec:discussion}

\begin{figure*}
    \centering
    \includegraphics[width=0.75\textwidth]{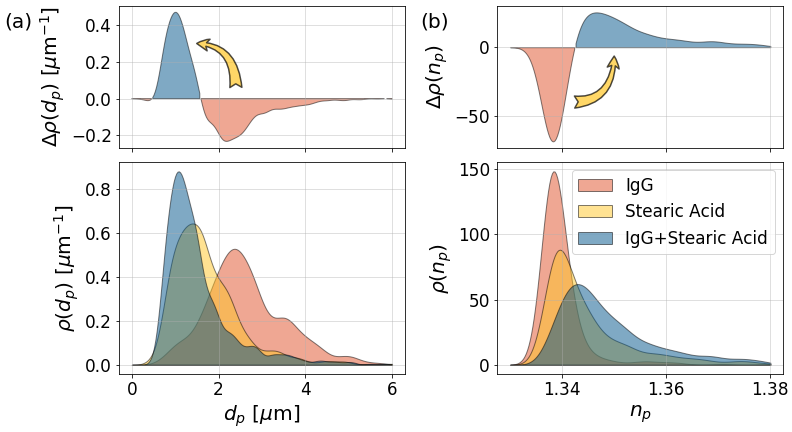}
    \caption{Interaction between stearic acid and IgG aggregates. (a) Mixing a dispersion of 
    stearic acid particles
    with a dispersion of IgG aggregates shifts the
    distribution of particle diameters,
    $\rho(d_p)$, to smaller sizes.
    This shift is apparent in the
    difference, $\Delta \rho(d_p)$, between
    the diameter distribution in the mixed
    sample and the diameters of stearic acid
    particles in IgG aggregates individually.
    (b) The distribution of refractive indexes
    shifts upward upon mixing. Taken together,
    (a) and (b) show that the mixture of stearic
    acid and IgG favors aggregates that are smaller
    and denser than either IgG or stearic acid
    alone.}
    \label{fig:stearic}
\end{figure*}

Holographic video microscopy detects the wide variety
of subvisible particle types that can be present in
biologic pharmaceutical formulations and provides
a physical basis for differentiating the different species, 
quantifying their properties and measuring their concentrations.
The data presented here demonstrate that HVM yields
accurate results for NIST-traceable particle standards,
subvisible protein aggregates, silicone oil emulsion
droplets, air bubbles, metal particles and fatty
acids that model the breakdown products of common
surfactants.
HVM provides consistent results, furthermore, in fluid
media whose viscosities range from \SI{1}{\centi P}
to at least \SI{20}{\centi P}.

In cases such as the model system in Fig.~\ref{fig:3mix},
disparate populations of particles 
can coexist in a multicomponent dispersion without
influencing each others' properties.
The results from this kind of heterogeneous mixture is apparent from 
the HVM data in Figs.~\ref{fig:3mix}(c)
and \ref{fig:3mix}(d) because the distribution of
properties in
a three-component sample can be reconstituted as
a superposition of the three separate stock samples.

In others cases, the particles in a heterogeneous
dispersion interact, yielding measureable changes in the
sample's HVM signature.
The data in Fig.~\ref{fig:stearic} illustrate
such a change.
These data were acquired for IgG aggregates,
stearic acid particles, and a mixture of these
two types of particles, each dispersed in water.
Figure~\ref{fig:stearic}(a) presents the distribution of particle diameters,
$\rho(d_p)$, for each of these three samples together with the
difference, $\Delta \rho(d_p)$, between the size distribution of the
mixture and the combined size distribution for the two homogeneous components.
Figure~\ref{fig:stearic}(b) shows the corresponding results for the
distribution of refractive indexes, $\rho(n_p)$ and $\Delta \rho(n_p)$.

Stearic acid particles are not readily differentiated
from protein aggregates either by size or by
refractive index because the two populations
of particles have similar optical properties.
The mixed sample, however, has distinctly different
properties from either of the components.
The presence of stearic acid in a dispersion of IgG aggregates shifts
the size distribution toward smaller particles than were in either
parent population and simultaneously shifts the refractive index distribution
to higher values.
The observed transformation of the HVM signature
suggests that stearic acid may promote restructuring
of branched protein aggregates into denser, more compact forms.
No such transformation is evident in Fig.~\ref{fig:3mix} when
droplets of silicone and oleic acid are added to a dispersion of
IgG aggregates.
This distinction suggests that stearic acid may interact with protein
aggregates in a different manner than other codispersed
species,
and thus highlights the differing influences that may be exerted by
breakdown products of different surfactants used in biopharmaceuticals.

These complementary and contrasting 
examples demonstrates the new window that HVM 
provides into the microstructure and composition
of subvisible particles.
The ability to detect particles with widely varying
physical properties, to distinguish them by size
and composition and to measure their concentrations
provides valuable information that can be used
to diagnose problems
in formulation, manufacturing, 
distribution and storage of
biopharmaceutical products.

\section{Acknowledgments}
\label{sec:acknowledgments}

Research reported in this publication was supported
by the National Center For Advancing Translational
Sciences of the National Institutes of Health under Award Number R44TR001590. The content is
solely the responsibility of the authors and does not necessarily represent the official views of
the National Institutes of Health.

\section*{Appendix: Lorenz-Mie analysis
of holograms}

The electric field of the collimated laser that illuminates
the sample may be modeled as a linearly polarized plane wave
propagating along $\hat{z}$ with its polarization directed
along $\hat{x}$ in Cartesian coordinates:
\begin{equation}
    \label{eq:incidentbeam}
    \vec{E}_0(\vec{r}, t) = u_0 e^{i k z} e{-i \omega t} \, \hat{x},
\end{equation}
where $\omega$ is the laser's frequency and
$k = n_m \omega / c$ is its wavenumber in a medium
of refractive index $n_m$. Here, $c$ is the speed
of light in vacuum.
An illuminated particle at position $\vec{r}_p$ relative
to the center of the microscope's focal plane scatters
a portion of this field 
for its location to position $\vec{r}$
in the focal plane.
Provided the particle is not too much
larger than the wavelength of light,
this scattered field may be modeled as
\begin{equation}
    \label{eq:scatteredwave}
    \vec{E}_s(\vec{r}, t) 
    = 
    E_0(\vec{r_p}, t) \,
    \vec{f}_s(k (\vec{r} - \vec{r}_p),
\end{equation}
where $\vec{f}_s(k \vec{r})$ 
describes the light-scattering
properties of the particle.

The total field reaching position $\vec{r}$ in
the imaging plane is the superposition of the
incident and scattered fields.
The time-averaged
intensity recorded by a video camera 
therefore may be modeled as
\cite{lee07a}
\begin{equation}
    \label{eq:intensity}
    I(\vec{r})
    = 
    u_0^2 \, 
    \abs{\hat{x} + 
    e^{-i k z_p} \, \vec{f}_s( k (\vec{r}_p - \vec{r}))}^2
    + I_0,
\end{equation}
where $I_0$ is the calibrated dark count of the camera.
Equation~\eqref{eq:intensity} defines
the imaging plane to be at axial position
$z = 0$ and assumes that the particle is
upstream of that plane by axial displacement $z_p$.

We analyze single-particle holograms, such as the examples
in Fig.~\ref{fig:schematic} by fitting the recorded image pixel-by-pixel
to Eq.~\eqref{eq:intensity}.
This fitting procedure requires
an expression
the scattering function, $\vec{f}_s(k\vec{r})$,
that appropriately models the scattering particle's
geometry and composition.
The Lorenz-Mie theory of light scattering
\cite{bohren83,mishchenko02}
expresses
this function as an expansion,
\begin{equation}
    \label{eq:lorenzmieseries}
    \vec{f}_s(k \vec{r})
    =
    \sum_{n=1}^\infty
    i^n
    \frac{2n + 1}{n(n+1)}
    \left[
    i a_n \,
    \vec{N}^{(3)}_{e1n}(k \vec{r})
    -
    b_n \,
    \vec{M}^{(3)}_{o1n}(k \vec{r})
    \right],
\end{equation}
in the vector spherical harmonics,
\begin{widetext}
\begin{subequations}
  \label{eq:vectorsphericalharmonics}
\begin{align}
    \vec{M}^{(3)}_{o1n}(k \vec{r})
    & =
    \frac{\cos\phi}{\sin\theta} 
    P_n^1(\cos\theta) \, j_n(kr) \, \hat{\theta}
    - \sin\phi \frac{d P_n^1(\cos\theta)}{d \theta} \,
    j_n(kr) \, \hat{\phi} \\
    \vec{N}^{(3)}_{e1n}(k \vec{r})
    & =
    n(n+1) \cos\phi \, P_n^1(\cos\theta) 
    \frac{j_n(kr)}{kr} \, \hat{r}
    +
    \cos\phi \frac{d P_n^1(\cos\theta)}{d\theta}
    \frac{1}{kr}
    \frac{d}{dr} [r \, j_n(kr)] \, \hat{\theta}
    - 
    \frac{\sin\phi}{\sin\theta}
    P_n^1(\cos\theta) \frac{1}{kr}
    \frac{d}{dr} [r \, j_n(kr)] \, \hat{\phi},
\end{align}
\end{subequations}
where $P_n^1(\cos\theta)$
is an associated Legendre polynomial and
$j_n(kr)$ is a spherical Bessel function of the first kind.
The vector spherical harmonics provide the natural basis
for solutions of Maxwell's wave equation in spherical
polar coordinates, $\vec{r} = (r, \theta, \phi)$.

The characteristics of the scattering particle are
encoded in the Lorenz-Mie scattering
coefficients, $a_n$ and $b_n$.
For the particular case of 
a homogeneous isotropic sphere
of radius $a_p = d_p/2$ 
and refractive index $n_p$,
these coefficients are
\cite{bohren83}
\begin{subequations}
\label{eq:expansioncoefficients}
\begin{align}
    a_n
    & =
    \frac{ 
    m^2 j_n(m k a_p) \, [k a_p \, j_n(k a_p)]'
    -
    j_n(k a_p) \, [m k a_p \, j_n(m k a_p)]'
    }{
    m^2 j_n(m k a_p) \, [k a_p h_n^{(1)}(k a_p)]'
    -
    h_n^{(1)}(k a_p) \, [m k a_p \, j_n(m k a_p)]'
    } \\
    b_n
    & =
    \frac{ 
    j_n(m k a_p) \, [k a_p \, j_n(k a_p)]'
    -
    j_n(k a_p) \, [m k a_p \, j_n(m k a_p)]'
    }{ 
    j_n(m k a_p) \, [k a_p h_n^{(1)}(k a_p)]'
    -
    h_n^{(1)}(k a_p) \, [m k a_p \, j_n(m k a_p)]'
    } ,
\end{align}
\end{subequations}
\end{widetext}
where $h_n^{(1)}(kr)$ is a spherical Hankel function of
the first kind, $m = n_p/n_m$ is the refractive
index contrast between the particle and the surrounding
medium, and primes
represent derivatives with respect to the argument.

Each nonlinear least-squares fit of a 
recorded hologram to this generative model
involves finding
values for particle's three-dimensional 
position, $\vec{r}_p$, 
diameter, $d_p$, and refractive index, $n_p$,
that minimize residuals between the
hologram and the prediction of 
Eq.~\eqref{eq:intensity}.
The software implementation of 
Eq.~\eqref{eq:intensity}
through Eq.~\eqref{eq:expansioncoefficients}
in the xSight used for the present study 
typically converges
to a solution with uncertainties in 
position of
$\Delta x_p = \Delta y_p \leq \SI{1}{\nm}$ and
$\Delta z_p \leq \SI{3}{\nm}$,
uncertainty in diameter of
$\Delta d_p \leq \SI{5}{\nm}$
and uncertainty in refractive index of
$\Delta n_p \leq \num{3e-3}$.
Each single-particle fit requires
roughly \SI{50}{\milli\second}
and each particle is recorded and
analyzed up to \num{10} times to ensure
reliable characterization results.
A representative sample of
a few thousand particles therefore
can be analyzed in ten minutes or so.

%\bibliographystyle{publist}
%\bibliography{differentiation}

\end{document}